\documentclass[iop,apj]{emulateapj}

\shortauthors{R. FUSCO-FEMIANO \& A. LAPI}
\shorttitle{SM ANALYSIS OF A1246 AND J255}
\slugcomment{Accepted by ApJ}

\begin{document}

\title{SuperModel Analysis of Abell 1246 and J255: \\on the Evolution of Galaxy Clusters from High to Low Entropy States}
\author{R. Fusco-Femiano\altaffilmark{1}, A. Lapi\altaffilmark{2,3,4,5}}
\altaffiltext{1}{IAPS-INAF, Via Fosso del Cavaliere, 00133 Roma, Italy -
roberto.fuscofemiano@iaps.inaf.it}\altaffiltext{2}{Dip. Fisica, Univ. `Tor
Vergata', Via Ricerca Scientifica 1, 00133 Roma,
Italy.}\altaffiltext{3}{SISSA, Via Bonomea 265, 34136 Trieste,
Italy}\altaffiltext{4}{INFN - Sezione di Trieste, Via Valerio 2, 34127
Trieste, Italy }\altaffiltext{5}{INAF - Osservatorio Astronomico di Trieste,
via Tiepolo 11, 34131 Trieste, Italy}

\begin{abstract}
We present an analysis of high-quality X-ray data out to the virial radius
for the two galaxy clusters Abell 1246 and GMBCG J255.34805+64.23661 (J255)
by means of our entropy-based SuperModel. For Abell 1246 we find that the
spherically-averaged entropy profile of the intracluster medium (ICM)
progressively flattens outwards, and that a nonthermal pressure component
amounting to $\approx 20\%$ of the total is required to support hydrostatic
equilibrium in the outskirts; there we also estimate a modest value
$C\approx 1.6$ of the ICM clumping factor. These findings agree with
previous analyses on other cool-core, relaxed clusters, and lend further
support to the picture by Lapi et al. (2010) that relates the entropy
flattening, the development of nonthermal pressure component, and the
azimuthal variation of ICM properties to weakening boundary shocks. In this
scenario clusters are born in a high-entropy state throughout, and are
expected to develop on similar timescales a low entropy state both at the
center due to cooling, and in the outskirts due to weakening shocks.
However, the analysis of J255 testifies how such a typical evolutionary
course can be interrupted or even reversed by merging especially at
intermediate redshift, as predicted by Cavaliere et al. (2011b). In fact, a
merger has rejuvenated the ICM of this cluster at $z\approx 0.45$ by
reestablishing a high entropy state in the outskirts, while leaving intact
or erasing only partially the low-entropy, cool core at the center.
\end{abstract}


\section{Introduction}

\setcounter{footnote}{0}

The low density $n\la 10^{-4}$ cm$^{-3}$ of the intracluster medium (ICM) in
the outskirts of galaxy clusters has severely limited, until recently, the
capability of investigating those regions via the thermal bremsstrahlung
emission $L_{X}\propto n^2$ in X rays. On the other hand, cluster outskirts
are extremely interesting since: they provide the connection between the ICM
and the filamentary structures of the cosmic web; they are the regions where
most of the baryons and of the gravitationally dominant dark matter mass
reside; and they constitute the sites of several physical processes and
events affecting the ICM thermodynamic properties (see Kravtsov \& Borgani
2012; Cavaliere \& Lapi 2013; Reiprich et al. 2013). Thus nowadays the study
of cluster outskirts is a very hot topic, embracing both astrophysics and
cosmology.

A breakthrough in this field has been recently obtained with the advent of
the \textsl{Suzaku} X-ray observatory, thanks to its low and stable particle
background. The main, somewhat unexpected findings from the \textsl{Suzaku}
data taken so far can be summarized as follows.

\begin{itemize}

\item The ICM temperature rapidly declines outwards by a factor $3$ in the
    region $r\sim 0.3 - 1\,r_{200}$\footnote{Here $r_{\Delta}$ is the
    radius within which the mean density is $\Delta$ times the critical
    density, while $R$ is the viral radius of the cluster. Frequently used
    values read $r_{500}\approx R/2$ and $r_{200}\approx 3\, R/4$.} and
    slightly beyond (see Akamatsu et al. 2011; Reiprich et al. 2013); the
    temperature profiles are rather similar for relaxed and disturbed
    galaxy clusters (Kawarada et al. 2010; Ichikawa et al. 2013; Simionescu
    et al. 2013; Sato et al. 2014).

\item The entropy profile $k(r)$ flattens at $r\ga 0.5\, r_{200}$ (see
    Walker et al. 2012) relative to the shape $k \propto r^{1.1}$ expected
    from strong-shocked infall of external gas under pure gravitational
    infall (see Tozzi \& Norman 2001; Lapi et al. 2005; Voit 2005).

\item The thermodynamic properties of the ICM are subject to significant
    azimuthal variations (see Kawarada et al. 2010; Ichikawa et al. 2013;
    Sato et al. 2014). For some clusters it has been possible to ascertain
    that hot regions are adjacent to filamentary structures, while cold
    regions are in contact with low-density, void-like environments; this
    indicates a more efficient thermalization in the overdense infall
    regions, and highlights how the environment surrounding the cluster
    affects the physical processes in the outskirts.

\item In some relaxed cluster the mass profile derived from X-ray
    observations under the assumption of thermal hydrostatic equilibrium
    features an unphysical, decreasing behavior at large radii (e.g.,
    Kawaharada et al. 2010; Walker et al. 2012; Ichikawa et al. 2013; Okabe
    et al. 2014; Sato et al. 2014); this is a consequence of the rapid
    temperature decline, and can be explained in terms of an ICM far from
    thermal equilibrium, owing to the presence of a nonthermal pressure
    support (see Fusco-Femiano \& Lapi 2013, 2014). An example is the
    cluster Abell 1835 observed by \textsl{Chandra} (Bonamente et al. 2013)
    and by \textsl{Suzaku} (Ichikawa et al. 2013), where the underestimate
    of the hydrostatic mass implies a gas mass fraction $f_{\rm gas}$
    higher than the cosmic value at the virial radius, as also reported for
    several other clusters (Simionescu et al. 2011; Fusco-Femiano \& Lapi
    2013; 2014).

\end{itemize}

The actual existence of the entropy flattening revealed by the
\textit{Suzaku} observations was challenged by Eckert et al. (2013); these
authors estimated the entropy via the relation $k\propto p/n^{5/3}$ by
combining average \textit{ROSAT} gas density profiles (Eckert et al. 2012)
with average electron pressure profiles $p(r)$ from stacked
Sunyaev-Zel'dovich observations of $62$ clusters by the \textit{Planck}
satellite (see \textit{Planck} Collaboration et al. 2013). However,
Fusco-Femiano \& Lapi (2014) showed that such a procedure is not suited to
discriminate between a steady powerlaw increase and a flattening of the
entropy; this is because in hydrostatic equilibrium the dependence of $p(r)$
on $k(r)$ is much weaker than that of the temperature $T(r)$, so that the use
of X-ray temperature data is mandatory for precise determination of the
entropy profile in the outskirts.

In principle, the observed entropy flattening can be explained by gas
clumping; this causes an overestimate of the ICM density and hence an
underestimate of the entropy in the outskirts. On the other hand,
hydrodynamical simulations show that the clumping factor $C\equiv \langle
n^2\rangle/\langle n \rangle^2\la 2$ of the ICM in the outskirts is rather
limited and actually insufficient to explain the observed entropy flattening
(Mathiesen et al. 1999; Nagai \& Lau 2011; Vazza et al. 2013; Roncarelli et
al. 2013; Zhuravleva et al. 2013; Battaglia et al. 2014; Morandi \& Cui
2014), in agreement with the values for the clusters analyzed so far
(Fusco-Femiano \& Lapi 2013, 2014).

The presence of low-entropy gas in the outskirts can also be explained by
considering that ions and electrons thermalize downstream the boundary shock
on different timescales by Coulomb collisions (see Hoshino et al. 2010;
Akamatsu et al. 2011). However, as noted by Okabe et al. (2014), the
thermalization of the electrons can actually occur on the much shorter
timescale of wave-particle interactions via plasma kinetic instabilities. A
more recent proposal by Fujita et al. (2013) envisages that the entropy
generation at the boundary shocks is not complete because part of the infall
energy of the external gas is used to accelerate cosmic rays. We note that
both these effects would be stronger in dynamically active clusters like
Coma, where instead the entropy shows no clear evidences of flattening in
undisturbed regions, and is much higher (even when scaled by the different
mass) than in many relaxed clusters (Simionescu et al. 2013).

An alternative explanation of entropy flattening, steep temperature decline,
nonthermal motions, and azimuthal variations, advocates the weakening of the
boundary shocks (Lapi et al. 2010; Cavaliere et al. 2011b). In fact, weaker
shocks produce less entropy, are less efficient in thermalizing the infall
energy of the external gas, while allowing the residual one to seep inside
and originate nonthermal motions in the form of turbulence. The shock
weakening is in turn due to a reduced inflow, that can mainly occur under two
circumstances: (i) either in the late evolution of a cluster, when the gas is
accreted from the wings of the initial perturbations, and especially so at
low redshift when the cosmic acceleration sets in; (ii) or in a particular
sector of a cluster facing an underdense, void-like region.

In this scenario clusters are born in a high-entropy state throughout by the
strong shocks occurring at the time of formation. Then they are expected to
develop synchronously a low entropy state both at the center due to cooling
(possibly balanced by gains from bubbling or rekindled AGNs in the central
member galaxies, see Fabian 2012), and in the outskirts due to weakening
shocks.

In Cavaliere et al. (2011a) we have predicted that the typical evolutionary
course of galaxy clusters from high to low entropy states may be temporarily
interrupted or even definitely reversed by a major merger; this can
rejuvenate the ICM outskirts by adding entropy to reestablish the powerlaw
behavior expected from strong shocks, and may penetrate deep in the cluster
to partially erase the cool core. We shall see that the analysis of the
cluster GMBCG J255.34805+64.23661 (hereafter J255) demonstrates such an
instance to actually occur in nature.

As mentioned above, the rapid decline of the gas temperature in the outskirts
may cause the mass estimates based on X-ray observables and thermal
hydrostatic equilibrium to be biased low by a systematic $\approx 10\%-20\%$
relative to the determinations via strong and weak lensing measurements (see
Arnaud et al. 2007; Mahdavi et al. 2008, 2013; Lau et al. 2009; Battaglia et
al. 2013). Recently, Okabe et al. (2014) have conducted a multi-wavelength
analysis of $4$ relaxed clusters reporting an average hydrostatic-to-lensing
total mass ratio that decreases from $\approx 70\%$ to $\approx 40\%$ going
from $r_{500}$ to the virial radius. These values appear to be at variance
with numerical simulations that report a hydrostatic-to-true total mass ratio
of $\approx 80-90\%$ at the virial radius (Lau et al. 2013; Nelson et al.
2014).

We stress that the comparison between X-ray and weak lensing masses is
fundamental for understanding the ICM physical state. In particular, the
difference in the mass values obtained with these independent methods probes
the level of nonthermal pressure support needed to sustain hydrostatic
equilibrium. This additional pressure component may be due to turbulence
originated by several processes such as mergers, supersonic motions of
galaxies through the ICM, or infall of gas into the cluster from the
surrounding environment (see simulations by Nagai et al. 2007; Shaw et al.
2010; Burns et al. 2010; Vazza et al. 2011; Rasia et al. 2012). We remark
that such a nonthermal pressure support must be taken into account to improve
both our astrophysical understanding of cluster outskirts and the accuracy in
cluster mass determination for cluster cosmology (see Vikhlinin et al.
2009a).

The paper is organized as follows. In \S~2 we recall the formalism of our
entropy-based SuperModel (SM, Cavaliere et al. 2009), that allows a
self-consistent analysis of the X-ray observables in presence of nonthermal
pressure support. In \S~3 we exploit the SM to analyze Abell 1246 at $z =
0.19$ and J255 at $z = 0.45$ basing on the X-ray observations by
\textit{Suzaku} and \textit{Chandra}, respectively. In \S~4 we discuss the
results and draw our conclusions. Throughout the paper we adopt the standard
flat cosmology (Hinshaw et al. 2013; \textsl{Planck} Collaboration 2014) with
parameters in round numbers: $H_0 = 70$ km s$^{-1}$ Mpc$^{-1}$,
$\Omega_{\Lambda} = 0.73$, $\Omega_M = 0.27$. Then 1 arcmin corresponds to
$191$ kpc for Abell 1246, and to $349$ kpc for J255.

\section{SuperModel with Turbulence}

We briefly recall the entropy-based SM formalism in presence of a nonthermal
component. We write the total pressure $p_{\rm tot}(r) = p_{\rm th}(r) +
p_{\rm nth}(r) = p_{\rm th}(r)[1 + \delta(r)]$ in terms of the nonthermal to
thermal ratio $\delta(r) \equiv p_{\rm nth}/p_{\rm th}$. Expressing the
density in terms of temperature and entropy via $n\propto (T/k)^{3/2}$ and
using the equation of hydrostatic equilibrium yields the temperature profile
in the form
\begin{eqnarray}
\nonumber\frac{T(r)}{T_R} = \left[\frac{k(r)}{k_R}\right]^{3/5}\,
\left[\frac{1 + \delta_R}{1 + \delta(r)}\right]^{2/5}\,\left\{1 +
\frac{2}{5}\frac{b_R}{1+\delta_R}\right.\times\\
\\
\nonumber\times \left.\int_r^R {\frac{{\rm d}x}{x} \frac{v^2_c(x)}{v^2_R}\,
\left[\frac{k_R}{k(x)}\right]^{3/5}\, \left[\frac{1 + \delta_R}{1 +
\delta(x)}\right]^{3/5}}\right\}~;
\end{eqnarray}
here $v^2_c(r)\equiv G\, M(<r)/r$ is the squared circular velocity ($v_R^2$
is the value at the virial radius $R$), and $b_R$ is the ratio of $v^2_R$ to
the squared sound speed at $R$ (Cavaliere et al. 2009, 2011b).

For the spherically averaged entropy profile $k(r)$ we consider the basic
pattern $k(r) = k_c + (k_R - k_c)(r/R)^a$ including a central floor $k_c\sim
10-100$ keV cm$^2$ (or even $<10$ keV cm$^2$ for low-$z$ cool core clusters,
see Panagoulia et al. 2014) going into an outer powerlaw rise with slope
$a\sim 1$ out to the virial value $k_R\sim$ some $10^3$ keV cm$^2$ (see Voit
2005; Lapi et al. 2005). However, to model a possible flattening in the outer
region, we modify the profile beyond a break radius $r_b$ by allowing the
slope $a$ to change (Lapi et al. 2010); for the sake of simplicity, the
entropy slope is taken to decline linearly with a gradient $a^{\prime} \equiv
(a-a_R)/(R/r_b - 1)$. The quantities $k_c$, $a$, $r_b$ and $a^{\prime}$ are
free parameters to be determined with their uncertainties from fitting the
projected, spherically-averaged X-ray observables, taking into account
errorbars on both axes (i.e., radius and density/temperature); to this
purpose, we exploit the physics analysis tool of the function minimization
algorithm MINUIT\footnote{see
\texttt{http://seal.web.cern.ch/seal/snapshot/work-packages/
mathlibs/minuit/}}.

As to the profile of $\delta(r)$ we follow the prescription by Cavaliere et
al. (2011b) based on the classic theory of turbulence generation (see
Kolmogorov 1941; Monin \& Yaglom 1965; Inogamov \& Sunyaev 2003; Petrosian \&
East 2008; Brunetti \& Lazarian 2011) and on indications from hydrodynamical
simulations (Lau et al. 2009; Vazza et al. 2011). We adopt a shape $\delta(r)
= \delta_R\, e^{-(R-r)^2/\ell^2}$ decaying on the scale $\ell$ inward of a
round maximum $\delta_R$ at the boundary shock where the nonthermal component
originates. The quantities $\delta_R$ and $\ell$ are held fixed during the
fitting procedure.

The traditional equation to estimate the total gravitational mass $M(<r)$
within $r$ must be modified to take into account the nonthermal pressure
component; the outcome reads (Fusco-Femiano \& Lapi 2013)
\bigskip
\begin{eqnarray}
\nonumber M(<r) = - \frac{k_B \{T(r)\,[1 +\delta(r)]\}\,r^2}{\mu m_p\,G}\,
\left[\frac{1}{n_e(r)}\frac{d n_e(r)}{d r}\right. +\\
\\
\nonumber+\left.\frac{1}{T(r)}\frac{d T(r)}{d r} + \frac{\delta(r)}{1 + \delta(r)}
\frac{2}{l^2}(R - r)\right]~.
\end{eqnarray}
The ICM mass writes $M_{\rm gas} = 4\pi \mu_e m_p\int{\rm d}r~{n_e(r) r^2}$
where $\mu_e \approx 1.16$ is the mean molecular weight of the electrons.

\section{SuperModel Analysis for Abell 1246 and J255}

In this section we present the SM analysis of Abell 1246 and J255, by fitting
the spherically-averaged electron density and temperature profiles measured
by \textsl{Suzaku} (Sato et al. 2014) and \textsl{Chandra} (Wang \& Walker
2014), respectively.

\subsection{Abell 1246}

Abell 1246 is a cluster of galaxies at redshift $z=0.1902$ (NASA/IPAC
Extragalactic Database) that features a regular ICM distribution, as reported
by \textit{Suzaku} observations in the $0.5-5.0$ keV energy range (Sato et
al. 2014). The thermal emission is significantly detected out to $r_{200}$
and the temperature at this radius is a factor $\approx 3$ lower than at the
peak (see Fig.~1). We assume a virial radius $R = 2\, r_{500}$ with $r_{500}
\approx 6.1^{\prime}$ as derived by Sato et al. (2014) from the mass value
obtained under the assumption of thermal hydrostatic equilibrium. Note that
at $r_{500}$ thermal hydrostatic equilibrium holds to a very good accuracy;
this is also shown by Okabe et al. (2014), that in their joint
X-ray/weak-lensing study of $4$ relaxed clusters report ${\rm d}M/{\rm d}r <
0$ at $r\gtrsim 1.3\,r_{500}$, in agreement with previous analyses (e.g.,
Kawaharada et al. 2010; Ichikawa et al. 2013). A posteriori, we have also
checked that our derived virial mass is consistent with the adopted value of
$R$ in yielding a mean density $\approx 100$ times the critical one.

The azimuthal analysis by Sato et al. (2014) in four directions reports a
slightly lower temperature in the southeast sector. Assuming spherical
symmetry and a constant temperature in each annular region it is found that
the de-projected electron density of the northeast and southwest sectors
tends to be lower than that of the southeast and northwest sectors in the
radial range $r\approx (0.9 - 1.8)\,r_{500}$. The derived gravitational mass,
azimuthally-averaged, starts flattening and then decreasing beyond $r_{500}$,
thus indicating a break in the assumption of thermal hydrostatic equilibrium.
In particular, Sato et al. (2014) found a mass of $(4.3 \pm 0.4)\times
10^{14}\, M_{\odot}$ within $r_{500}$, consistent with that of $(3.9 \pm 0.1)
\times 10^{14}\, M_{\odot}$ obtained from \textit{Chandra} observations by
Vikhlinin et al. (2009b) through the $M_{500} - T_X$ scaling relation. In
Sato et al. (2014) the X-ray mass profile is compared with that derived from
the stacked weak lensing analysis by Okabe et al. (2010). The weak lensing
mass amounts to $\approx 5$ and $\approx 7.8\times 10^{14}\, M_{\odot}$
within $r_{500}$ and $r_{200}$, respectively. The X-ray and weak lensing
masses are consistent within $r_{500}$ but not at greater radii where the
former is appreciably lower than the latter. Relatedly, Sato et al. (2014)
found a gas mass fraction consistent with the cosmic baryon fraction at
$r_{500}$ but not at $r_{200}$; on the other hand, the weak lensing mass at
$r_{200}$ yields instead a baryon fraction in agreement with the cosmic
value.

The entropy profile of Abell 1246 is similar to that reported for all the
clusters observed out to the virial radius by \textit{Suzaku} (Bautz et al.
2009; Kawaharada et al. 2010; Hoshino et al. 2010; Akamatsu et al. 2011;
Walker et al. 2012; Ichikawa et al. 2013). The entropy increases with radius
up to $r\approx r_{500}$ and then flattens outwards. As reported also for
Abell 1835 and Abell 1689 (see Ichikawa et al. 2013), in Abell 1246 the
entropy flattening is more pronounced in some cluster sectors; in particular,
it is more evident in the southeast sector rather than in the northwest one,
that appears to face a filament. Besides, in this latter sector the
temperature is higher than those measured in other regions (as it happens in
A1689, see Kawaharada et al. 2010), whereas the electron densities are
consistent.

We fit the spherically-averaged X-ray temperature profile of Abell 1246 with
the SM finding that, within the measurement uncertainties, a flattening
entropy profile performs better than a simple powerlaw (see Fig.~1). The
corresponding SM fit to the gas density $n_e(r)$ is reported in the bottom
panel of Fig.~1. We obtain a central entropy value $k_c = 101\pm 15$ keV
cm$^2$ rather high for a relaxed cluster; this is likely due to the low
spatial resolution of \textsl{Suzaku}, which is insufficient to resolve the
expected presence of a cool core in the relaxed cluster Abell 1246. For the
other parameters we find $a = 0.95^{+0.22}$, $r_b = (0.53\pm 0.07)r_{500}$
and $a^{\prime} = -(1.22\pm 0.21)$.

We compute the total gravitational mass within $r$ (see blue line of Fig.~2)
from Eq.~2 by assuming thermal hydrostatic equilibrium ($\delta$ = 0). In
agreement with the analysis of Sato et al. (2014) the mass profile starts to
flatten at $r \approx r_{500}$ and then declines downward at $r\gtrsim 1.6\,
r_{500}$, yielding a gas mass fraction greater than the cosmic value at the
virial radius (blue line of Fig.~2). The decreasing mass profile clearly
indicates that the outskirts of Abell 1246 are not in thermal hydrostatic
equilibrium. We evaluate the level and radial shape of the nonthermal
component by requiring the X-ray mass to be in agreement (as suggested by
Sato et al. 2014) with the stacked weak lensing mass profile from Okabe et
al. (2013, updating Okabe et al. 2010). We find that the agreement is
recovered for $\delta_R = 0.3$ and $\ell = 0.5$, implying a nonthermal
pressure component $\approx 20\%$ of the total pressure at the virial radius.
We note that our SM analysis gives a mass value consistent at $r_{500}$ with
the revised estimate of Okabe et al. (2013), and with the value obtained by
Sato et al. (2014); this cross-checks the consistency in our adopted value of
the virial radius. Despite the presence of a nonthermal pressure component
that gives a higher total mass consistent with the weak lensing
determination, the gas mass fraction remains higher than the cosmic value
(see red line of Fig.~2), highlighting the presence of gas clumping with a
factor $C\approx 1.6$ at the virial radius.

Fig.~3 shows the entropy profiles obtained by the SM analysis, confronted
with that observed by \textit{Suzaku}. We again find that the flattening
profile matches better the one derived by Sato et al. (2014). With the above
value of clumping, the entropy is underestimated only by a factor
$C^{1/3}\approx 1.2$ at the virial radius, insufficient to explain the
observed entropy flattening. This implies that also for Abell 1246 the
entropy flattening in the outskirts is mainly related to the rapid decline of
the temperature, and not to an overestimate of the density caused by gas
clumping; this is in agreement with the results recently reported by
Fusco-Femiano \& Lapi (2014) for a sample of other $4$ clusters.

\subsection{J255}

To investigate the ICM in the cluster outskirts a feasible possibility is
given by deep \textit{Chandra} observations of distant clusters at redshift
$\gtrsim 0.2$. In particular, diffuse emission has been observed  out to the
virial radius in Abell 1835 (Bonamente et al. 2013), showing a sharp decline
of the temperature consistent with the \textit{Suzaku} observations (Ichikawa
et al. 2013).

The internal regions of J255 present an elongated X-ray structure most likely
due to a recent merger event, while the cluster appears relaxed at distance
$r\gtrsim 1^{\prime}$. This cluster shows a projection overlap with the X-ray
emission of Abell 2246 at $z = 0.23$, that appears evident only in a limited
sector (Wang \& Walker 2014). A cool core is also detected (see Fig.~4). The
conservative extent of the cluster is $\approx 1.1\,r_{200}$ where $r_{200} =
(4\pm 0.2)^{\prime}$ is estimated by the average temperature (5.5$\pm$0.4
keV) using the scaling relations by Arnaud et al. (2005); the total mass
$M_{200}$ amounts to $5.0^{+0.8}_{-0.7}\times 10^{14} M_{\odot}$.

We perform a fit with the SM to the de-projected temperature profile derived
by Wang \& Walker (2014), assuming spherical symmetry and a virial radius $R
= 4/3\,r_{200}$. Despite the central value of the outermost bin is limited to
$\sim r_{500}$, it is evident that the observed temperature profile does not
show the rapid decline observed in other clusters, and is well fitted by a
simple powerlaw increase of the entropy (see Fig.~4) $k \propto r^a$ with $a
= 0.98^{+0.2}$ (and $k_c = 12.5\pm 3.1$ keV cm$^2$), in agreement with the
expected value of $1.1$ from strong-shocked accretion under pure
gravitational infall. The corresponding fit to the gas density profile
observed by \textit{Chandra} is reported in Fig.~4.

The shape of the inferred entropy profile and the morphologically disturbed
features of J255 concur to suggest that the cluster has undergone a recent
merger event; on the other hand, the presence of a cool core implies that
either the merger has not reached the core yet, or it has only partially
destroyed the cool core (Fusco-Femiano et al. 2009; Rossetti \& Molendi
2010).

We find that the total cluster mass (see Fig.~5) from the SM is consistent
with the $M_{200}$ value estimated by Wang \& Walker (2014) and with a mean
density that at $R$ is $\approx 100$ times the critical one, so
cross-checking the consistency in our assumed value of the virial radius; the
overall mass corresponds to a gas mass fraction at $R$ slightly above the
cosmic value. This may be explained either by a gas clumping factor $C\approx
1.3$, or by a modest level of turbulence ($p_{\rm nth}\approx 5\% \, p_{\rm
tot}$).

\section{Discussion and conclusions}

Our SuperModel (SM) analysis of the X-ray temperature and brightness profiles
can shed light on the dynamical state of the intracluster medium (ICM);
specifically, it enables us to determine the shape of the entropy profile
throughout the cluster volume, from the inner core region out to the
outskirts. In the latter, our analysis provides both a measure of the
nonthermal pressure support needed to sustain hydrostatic equilibrium, and an
estimate of gas clumping. In this paper we have applied the SM analysis to
the two clusters Abell 1246 at $z = 0.19$ and J255 at $z = 0.45$, that have
been recently observed out to the virial radius by \textit{Suzaku} and
\textit{Chandra}, respectively.

We have determined the entropy profile of Abell 1246 by fitting with the SM
the \textit{Suzaku} temperature data (see Fig.~1); we have found that the
entropy progressively flattens outwards relative to the simple powerlaw
increase $k\propto r^{1.1}$ expected from strong-shocked accretion under pure
gravitational infall; this is accord with the entropy analysis by Sato et al.
(2014).

We have determined the X-ray mass on assuming pure thermal hydrostatic
equilibrium, finding result consistent with the revised stacked cluster
lensing measurements by Okabe et al. (2013) at $r_{500}$. On the other hand,
the determination at the virial radius is biased low due to the break of
thermal hydrostatic equilibrium, related to the rapid decrease of the
temperature as also found by Okabe et al. (2014); quantitatively, we derive a
bias of $\approx 20\%$ consistent with the X-ray to true total mass ratio
within $r_{200}$ found by hydrodynamical simulations (see Lau et al. 2013;
Nelson et al. 2014), but appreciably lower than the result obtained by Okabe
et al. (2014) using a joint X-ray and weak lensing analysis for a sample of
$4$ relaxed clusters.

We have confirmed previous analysis (Sato et al. 2014) in finding that the
mass profile reconstructed from X-ray observables features in the outskirts
an unphysical, non-monotonic behavior. This is likely due to the break of
thermal hydrostatic equilibrium in the outskirts. Thus we have exploited the
capability of our SM to include a nonthermal pressure component. We have
quantified such nonthermal levels by requiring the reconstructed mass to be
consistent with the weak lensing determination (see Fig.~2); we find a
nonthermal pressure component of about $20\%$ of the total at the virial
radius.

We have also estimated the level of gas clumping by matching the resulting
gas mass fraction to the cosmic value; we find a clumping factor $C\approx
1.6$ at the virial radius. Modest values of $C$ are consistent with our
previous analysis of relaxed clusters (see Fusco-Femiano \& Lapi 2013, 2014)
mainly based on gas density profile measured by \textsl{ROSAT}, which by its
coarser resolution is less sensitive to clumping. Our findings also agree
with the bounds $C\la 2$ at the virial radius from numerical simulations (see
Nagai \& Lau 2011; Vazza et al. 2013; Battaglia et al. 2014). Moreover, the
modest levels of gas clumping imply that the entropy at the virial radius may
be underestimated only by a factor $C^{1/3}\approx 1.2$, in agreement with
our previous analysis of relaxed clusters and with the more recent study by
Okabe et al. (2014). This indicates that the entropy flattening in the
outskirts is strictly related to the steeply declining temperature profile
and not to an overestimate of the density because of clumping.

The shape of the entropy profile strongly suggests that Abell 1246 is a
relaxed cluster that probably had time to develop a cool core in the inner
region, though it cannot be resolved because of the insufficient spatial
resolution of \textit{Suzaku}. The indication of the relaxed nature of Abell
1246 is confirmed by the physical correlation between the outskirts entropy
and the virial mass suggested by Walker et al. (2012), Sato et al. (2012) and
recently investigated by Okabe et al. (2014) using the joint X-ray/weak
lensing analysis of $4$ relaxed clusters. Specifically, this correlation is
between the average entropy $K_{\rm out}$ in the range $r_{500} - R$, and
$M_{\rm vir}\,E(z)$, with
$E(z)=[\Omega_{m,0}\,(1+z)^3+\Omega_{\Lambda}]^{1/2}$ the Hubble expansion
rate. For Abell 1246 we found an average entropy $K_{\rm out}$ of $\approx
1300$ keV cm$^2$ at $M_{\rm vir}\, E(z)\,[10^{14}\, M_{\odot}]\approx 10$
consistent with the tight correlation found by Okabe et al. (2014) for
relaxed clusters.

This implies that the actual weak lensing mass profile of Abell 1246 is not
much different from that adopted by us on the basis of Okabe et al. (2013,
updating Okabe et al. 2010); thus our derived values of $\delta_R$ and $C$
are not significantly affected. The azimuthal analysis performed by Sato et
al. (2014) reveals the entropy flattening to be more pronounced in the
cluster sectors facing low-density, void-like environments, while to be
almost absent in the sectors facing high-density filaments of the cosmic web.

These findings agree with previous analyses on other cool-core, relaxed
clusters, and lend further support to the picture by Lapi et al. (2010) and
Cavaliere et al. (2011a) that relates the entropy flattening and the
development of nonthermal pressure component to weakening boundary shocks;
the latter produce less entropy, while allowing more bulk inflow energy to
seep inside and develop nonthermal pressure in the form of turbulence. The
shock weakening mainly occurs under two conditions: (i) either at late time
in cluster evolution when external gas is accreted from the wings of the
initial perturbation, and especially so at low redshift in an accelerating
background Universe; (ii) or in a particular sector of a cluster facing an
underdense, void-like region. In this scenario clusters are born in a
high-entropy state throughout, and are expected to develop synchronously a
low entropy state both at the center due to cooling, and in the outskirts due
to weakening shocks.

However, the analysis of J255 testifies how such a typical evolutionary
course can be interrupted or even reversed by merging, especially at
intermediate redshift. We have determined the entropy profile by fitting with
SM the azimuthally average temperature profile observed by \textit{Chandra}.
We have found that the entropy steadily increase with radius as a powerlaw
$k\propto r^{1.1}$, close to the expectation for strong shocks (see Fig.~4).

We have also determined the mass profile (see Fig.~5), finding a value of
$M_{200}\approx 5.6\pm 0.4\times 10^{14}\, M_\odot$ in agreement with that
derived by Wang \& Walker (2014) using the mass-temperature scaling relation
(Arnaud et al. 2005). The gas mass fraction $f_{\rm gas}$ at the virial
radius is slightly greater than the cosmic value (see Fig.~5). This can be
easily explained by a clumping factor $C\approx 1.3$ or by a very modest
level around $5\%$ of nonthermal pressure; the latter would imply a total
mass in excess of $\approx 7\%$ than the value reported in Fig.~5 at $R$.

The value of the average outer entropy $K_{\rm out}\approx 2140$ keV cm$^2$
at $M_{\rm vir}\, E(z)\, [10^{14} M_{\odot}]\approx$ 10 is much larger than
the value derived for relaxed clusters; this concurs with the powerlaw shape
of the entropy profile in indicating that J255 as an unrelaxed cluster. This
is also confirmed by the interesting morphological structure of J255. The
cluster features a cool core, but with an elongated X-ray morphology in the
inner regions, most likely due to a recent merger event.

So the typical evolutionary course from high to low entropy state expected to
occur simultaneously both in the core due to cooling and in the outskirts by
reduced entropy production in weakening shocks has been interrupted by a
merger (see Cavaliere et al. 2011a; Cavaliere \& Lapi 2014). This event may
have indeed reheated locally the ICM, rejuvenating the cluster outskirts to a
high entropy state but leaving intact or destroying only partially the
low-entropy cool core (Fusco-Femiano et al. 2009; Rossetti \& Molendi 2009).
In fact, such a behavior is in pleasing agreement with the predictions by
Cavaliere et al. (2011a).

Finally, we stress that the analysis of these two clusters have yielded a
further evidence that the entropy content in the outskirts is jointly
determined by the dynamical state of the cluster and by the surrounding
environment.

\begin{acknowledgements}
We thank our referee for constructive comments. This work has been supported
in part by the MIUR PRIN 2010/2011 `The dark Universe and the cosmic
evolution of baryons: from current surveys to Euclid', by the INAF PRIN
2012/2013 `Looking into the dust-obscured phase of galaxy formation through
cosmic zoom lenses in the Herschel Astrophysical Terahertz Large Area
Survey'. A.L. thanks SISSA for warm hospitality.
\end{acknowledgements}

\clearpage
\begin{figure*}
\begin{center}
\epsscale{1.15}\plottwo{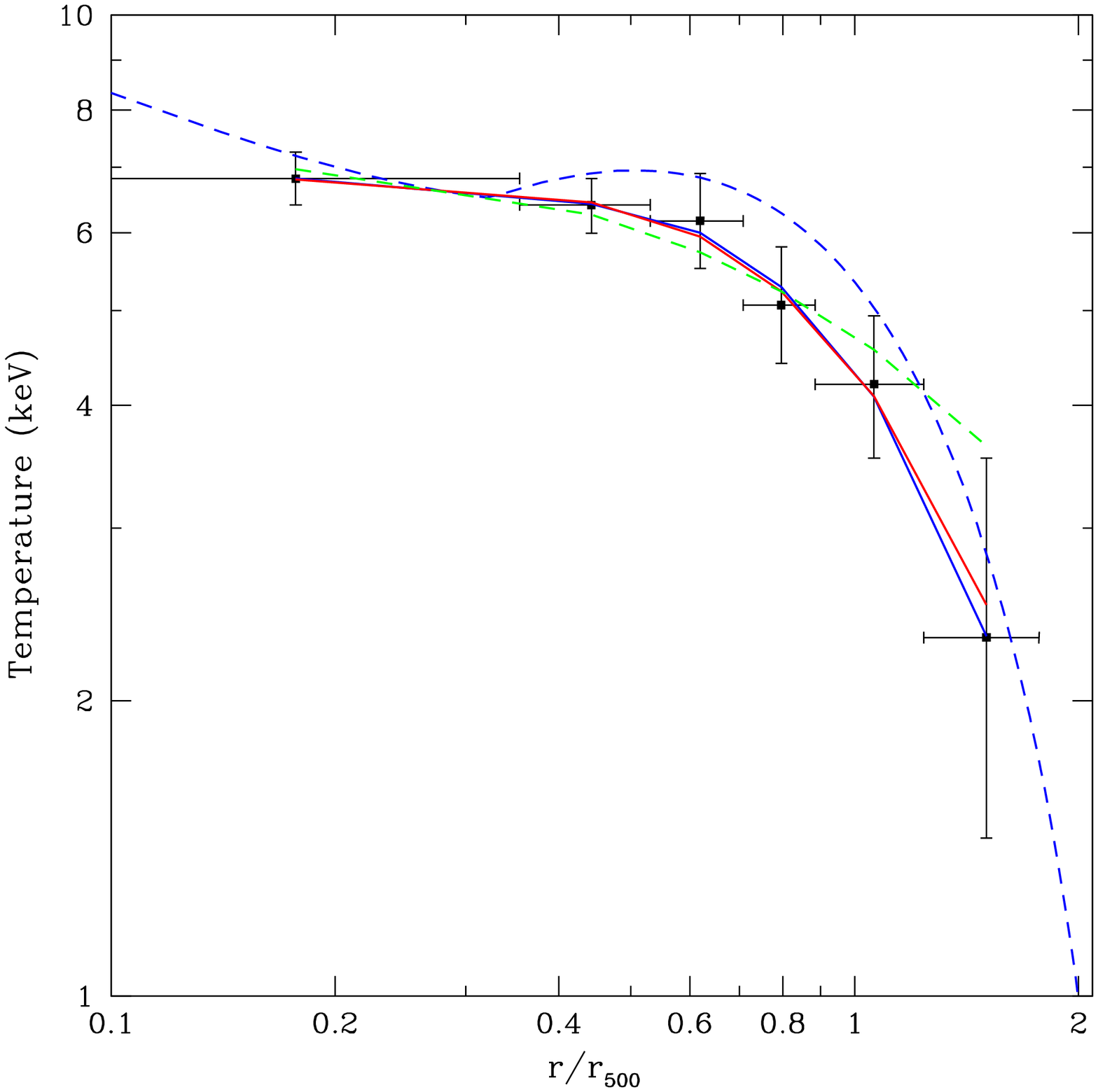}{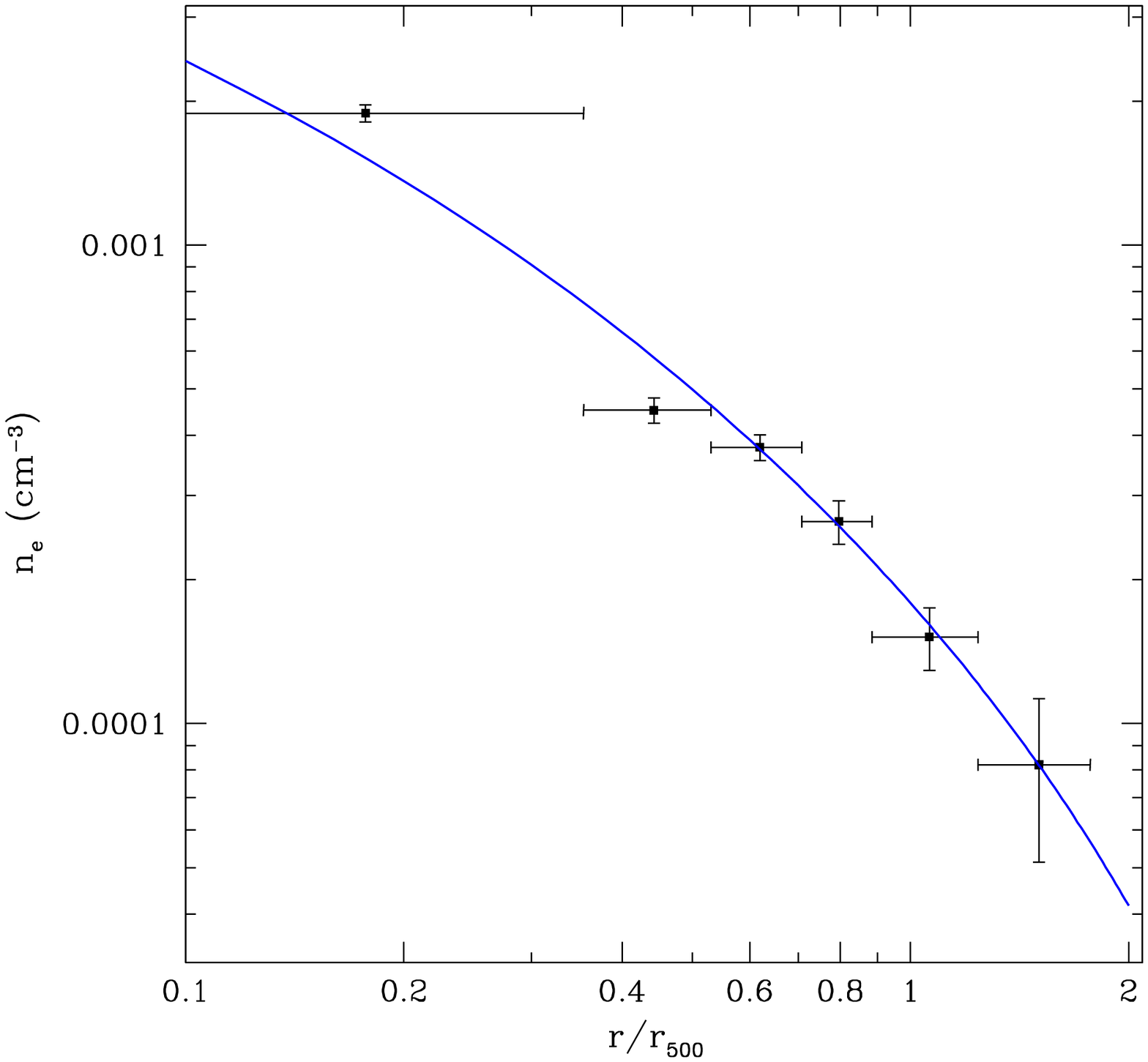}\bigskip\bigskip\bigskip\bigskip\caption{Abell
1246 - Top panel: Projected temperature profiles. Data points from
\textsl{Suzaku} (Sato et al. 2014). Blue and red lines illustrate the SM fits
without ($\delta_R$ = 0) and with nonthermal pressure component ($\delta_R =
0.3$, $\ell = 0.5$; see \S~2 for details), respectively. Both lines are
obtained with a flattening entropy profile, while the the dashed green line
is obtained with a simple powerlaw shape; dashed blue line shows the
deprojected temperature profile corresponding to the blue line. Bottom panel:
The blue line illustrates the SM fit to the electron density data points from
\textsl{Suzaku} (Sato et al. 2014). We assume $r_{500} = 6.1^{\prime}$ as
estimated by Sato et al. (2014) under the assumption of thermal hydrostatic
equilibrium (see discussion in Sect. 3.1).}
\end{center}
\end{figure*}

\clearpage
\begin{figure*}
\begin{center}
\epsscale{1.15}\plottwo{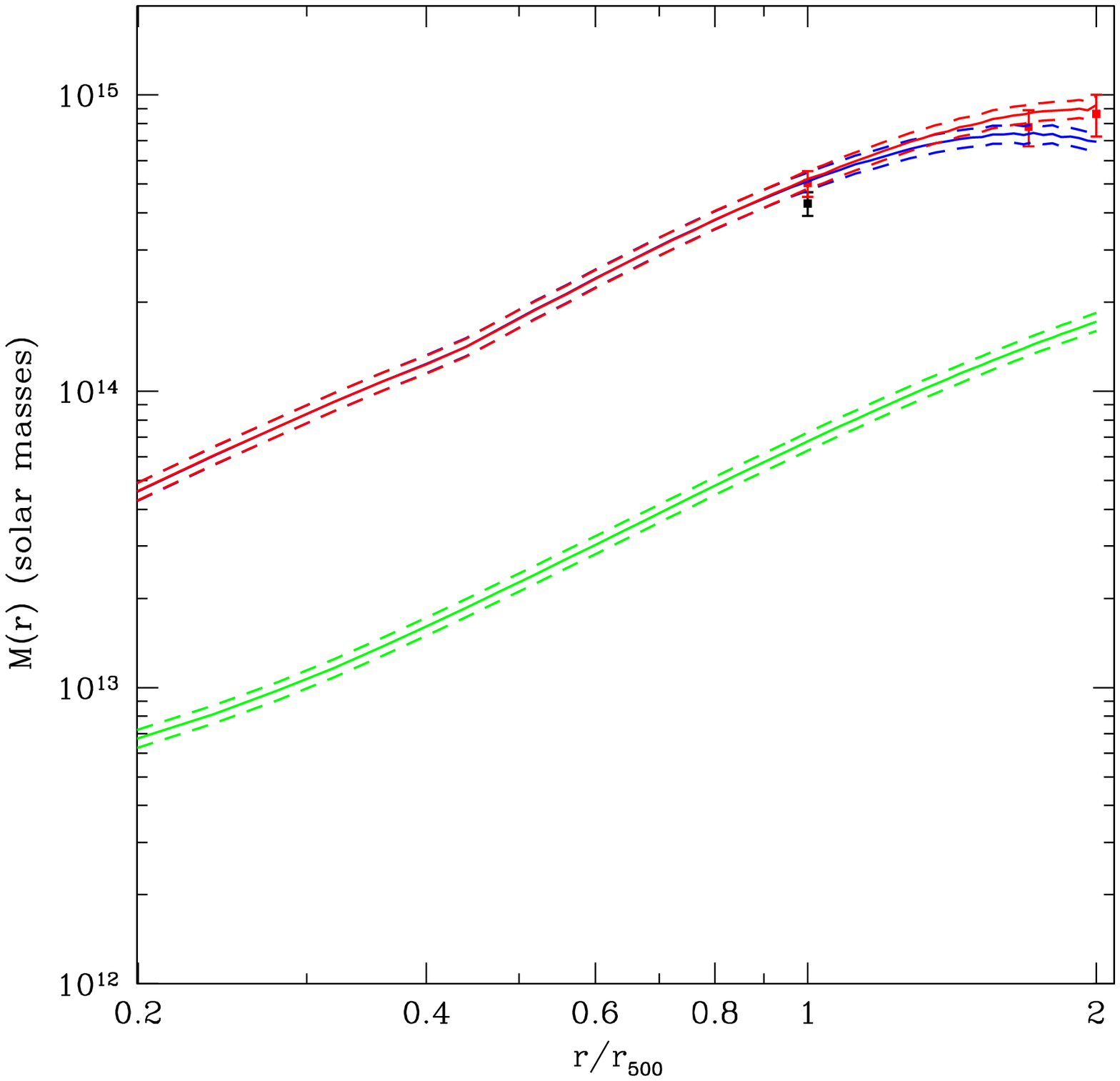}{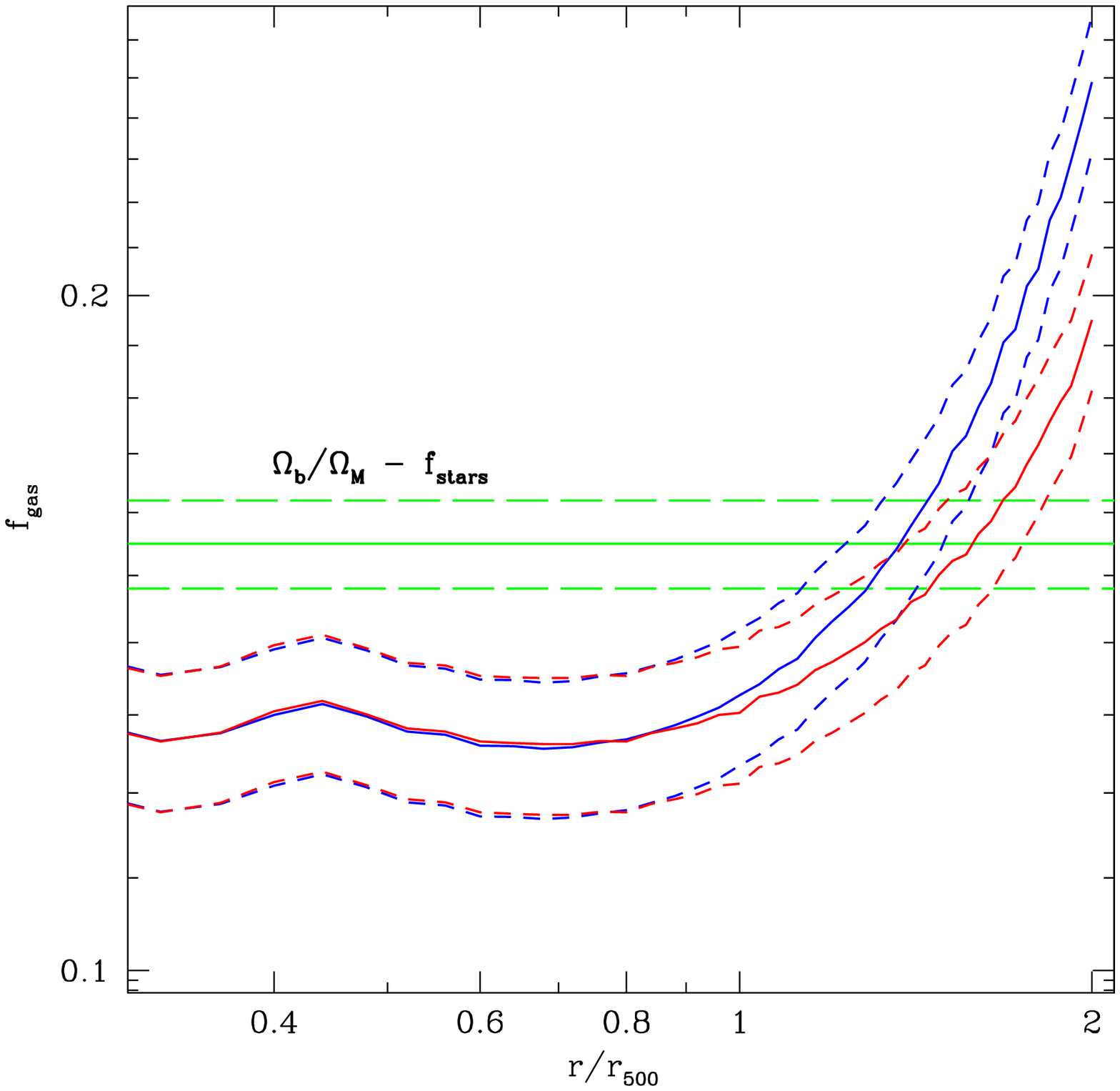}\bigskip\bigskip\bigskip\bigskip\caption{Abell
1246 - Top panel: Mass profiles. Blue line illustrates the X-ray cluster mass
obtained with $\delta_R = 0$, while the red line is obtained with $\delta_R =
0.3$ and $\ell = 0.5$; green line illustrates the gas mass obtained by the
gas density from \textsl{Suzaku} (Sato et al. 2014). The dashed lines mark
the $1\sigma$ uncertainty region from the SM fit. The black point shows the
mass derived at $r_{500}$ by Sato et al. (2014), while the red points refer
to the weak lensing mass values from the stacking analysis by Okabe et al.
(2013, updating Okabe et al. 2010). Bottom panel: Gas mass fraction
derived from the above mass profiles. Blue line is with $\delta_R = 0$; red
line is with the above values of $\delta_R$ and $\ell$; green lines show the
measured difference of the cosmic baryon fraction and the fraction of baryons
in stars and galaxies, $\Omega_b/\Omega_M -f_{\rm stars} = 0.155 \pm 0.007$
(Komatsu et al. 2011; Gonzalez et al. 2007).}
\end{center}
\end{figure*}

\clearpage
\begin{figure*}
\begin{center}
\epsscale{0.516}\plotone{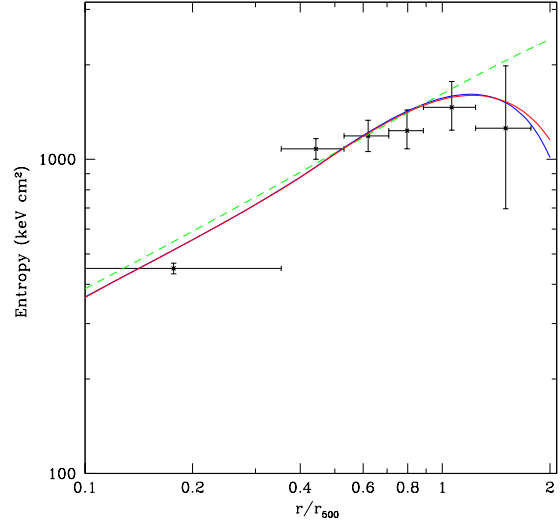}\bigskip\bigskip\bigskip\bigskip\caption{Abell 1246 - Entropy
profiles. Blue line illustrates the entropy profile obtained with the deprojected
temperature profile (dashed blue line) and the gas density profile in
Fig.~1. Red line shows the entropy profile obtained with the deprojected
temperature profile relative to $\delta_R = 0.3$ and $\ell = 0.5$
(see red line of Fig.~1). Dashed green line is obtained by the fit to the
temperature data with a powerlaw increase of the entropy. Data points show the
entropy values obtained from the analysis by Sato et al. (2014).}
\end{center}
\end{figure*}

\clearpage
\begin{figure*}
\begin{center}
\epsscale{1.15}\plottwo{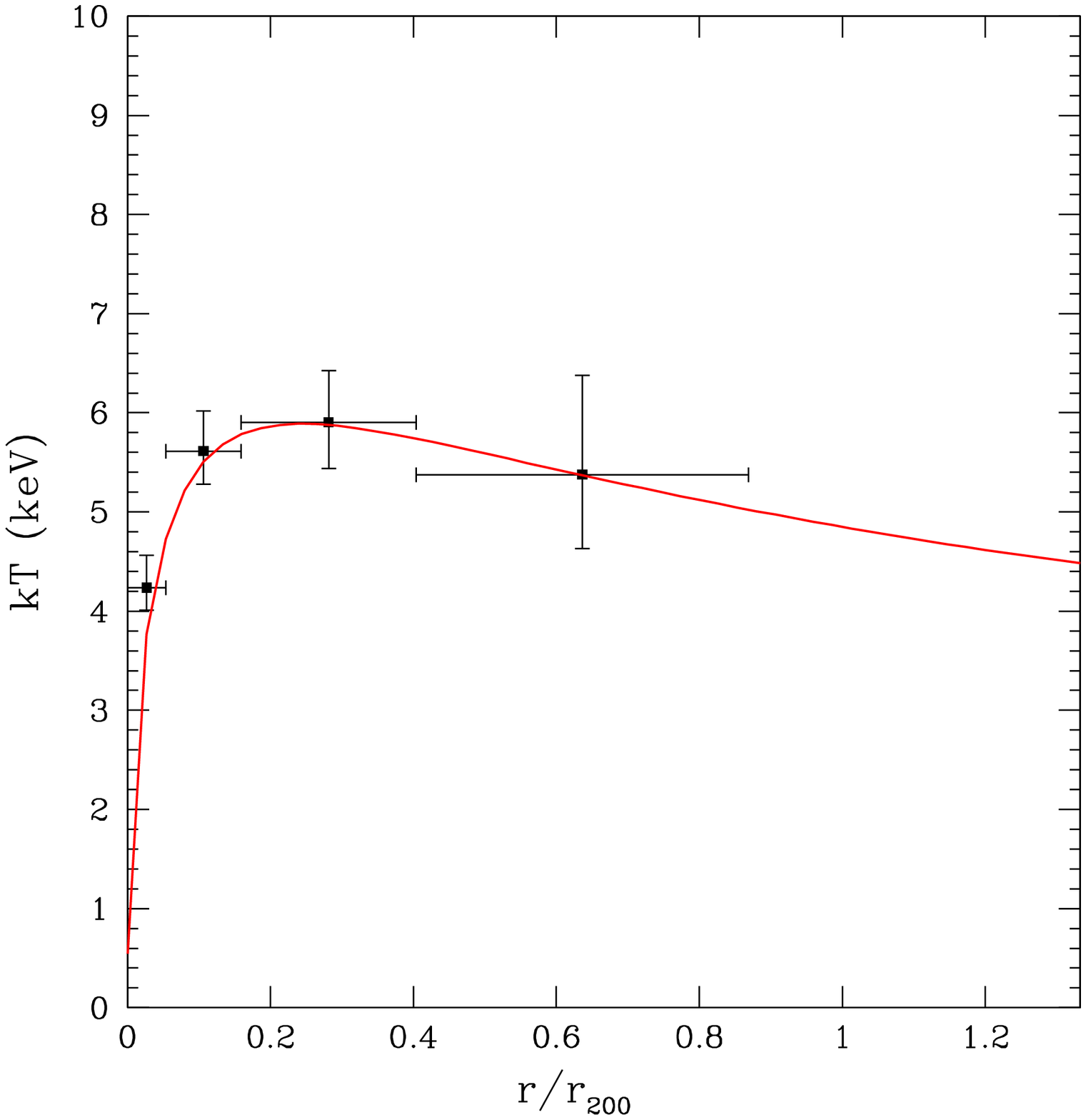}{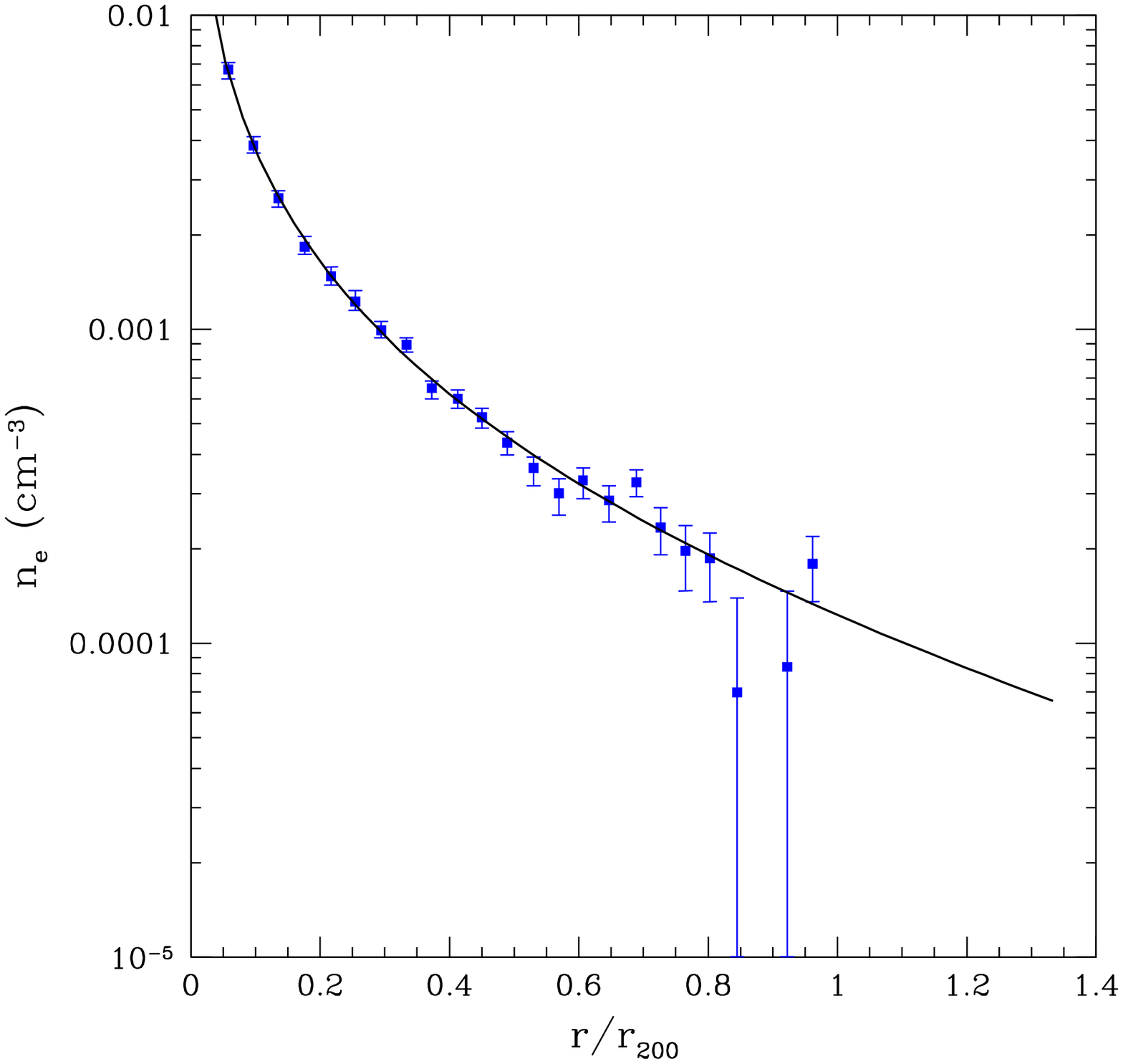}\bigskip\bigskip\bigskip\bigskip\caption{J255 - Top panel:
Temperature profile. The deprojected data points are derived by Chandra
observations (Wang \& Walker 2014). The red line illustrates the fit obtained with an
entropy profile following a simple powerlaw increase; this fit is
indistinguishable from that obtained with entropy flattening. Bottom panel:
black line shows the gas density profile obtained by the SM fit to the points
derived by Chandra observations (Wang \& Walker 2014). We assume
$r_{200} = 4^{\prime}$ as estimated by Wang \& Walker (2014) using the
scaling relations by Arnaud et al. (2005).}
\end{center}
\end{figure*}

\clearpage
\begin{figure*}
\begin{center}
\epsscale{1.15}\plottwo{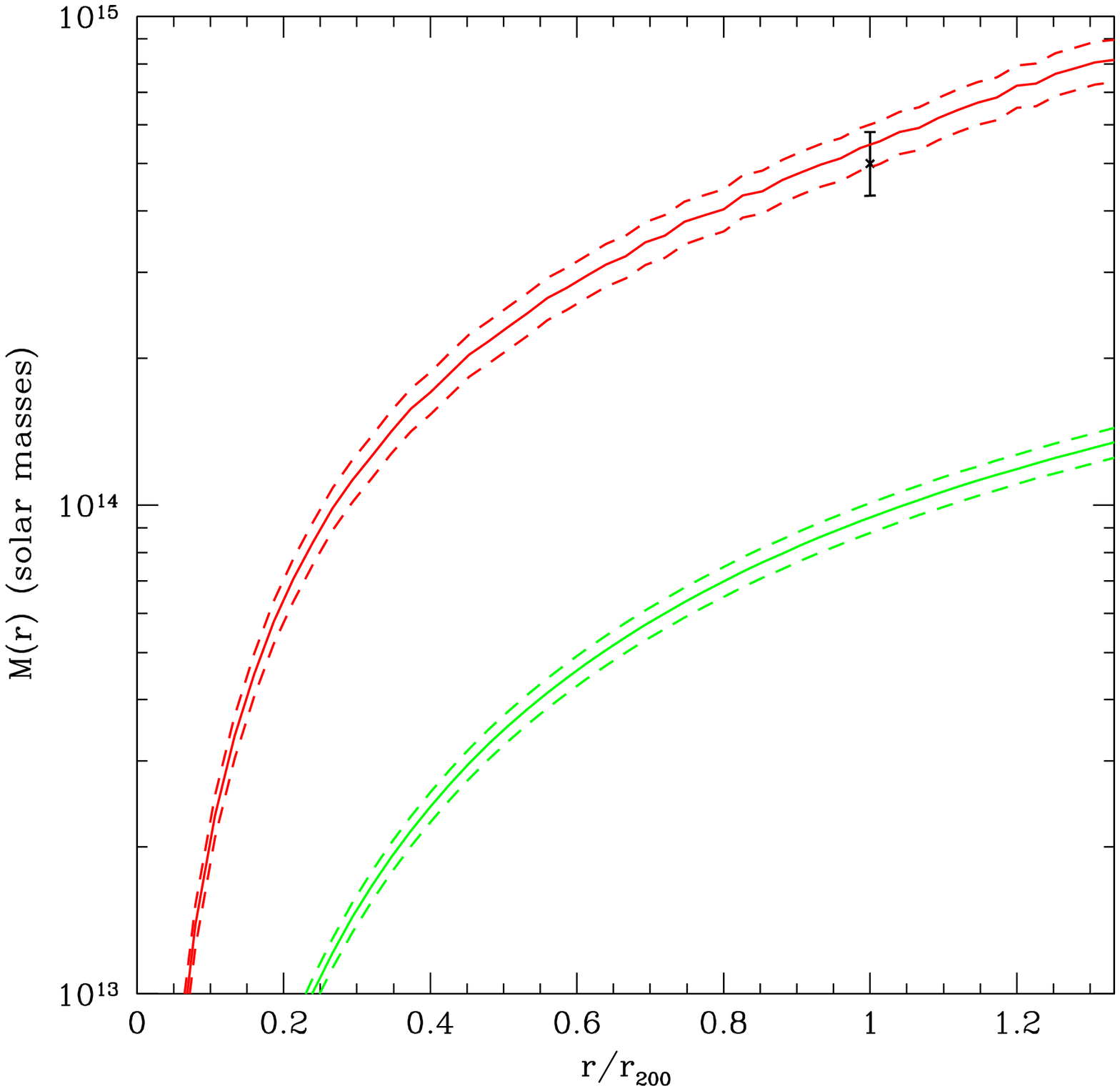}{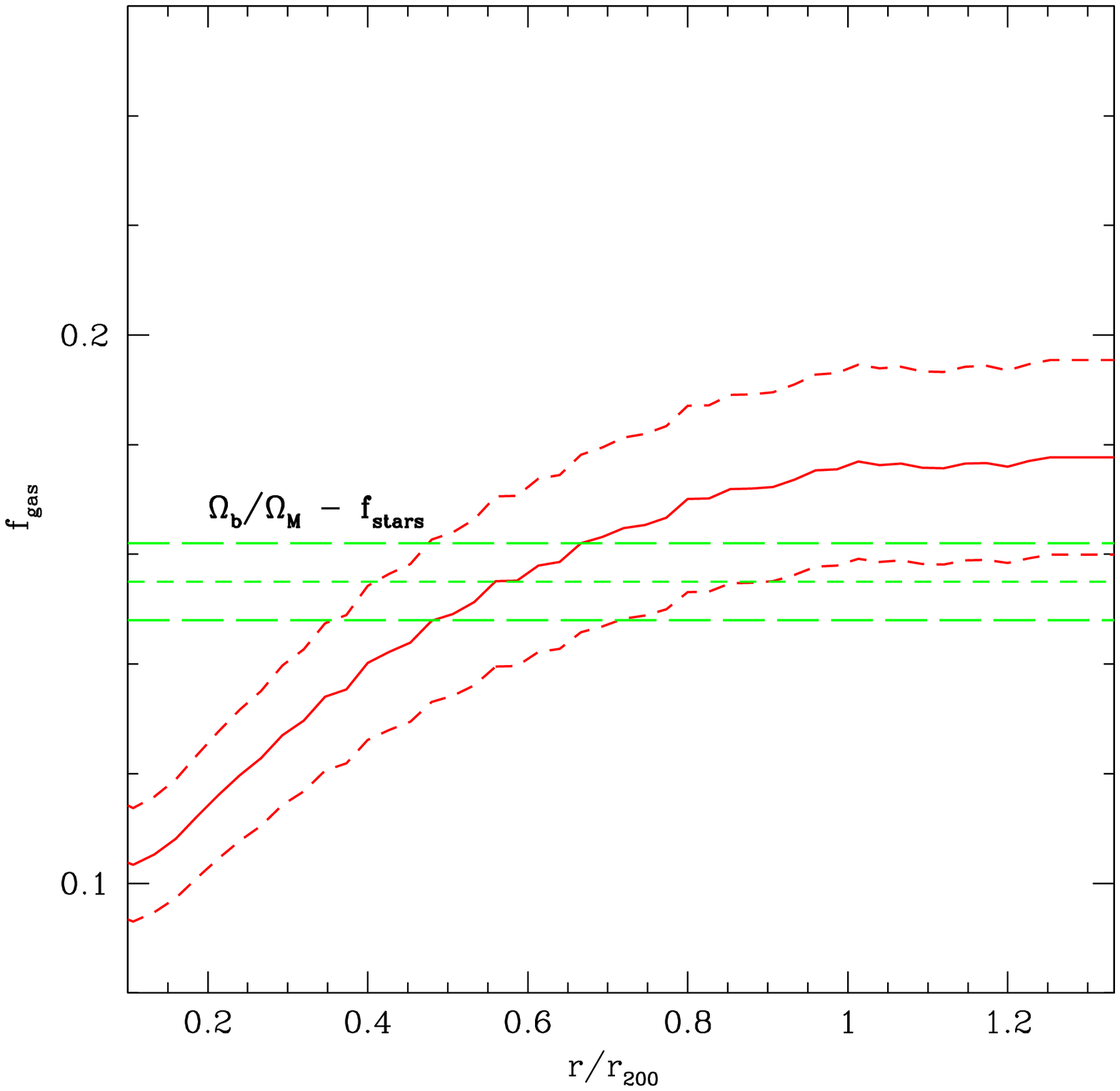}\bigskip\bigskip\bigskip\bigskip\caption{J255
- Top panel: Mass profiles. The red line illustrates the mass profile
obtained with the gas temperature and density profiles of Fig.~4; the data
point is the value of $M_{200}$ obtained by Wang \& Walker (2014) using
scaling relations with the temperature by Arnaud et al. (2005); the green
line shows the gas mass profile. The dashed lines mark the $1\sigma$
uncertainty region from the SM fit. Bottom panel: the red curve is obtained
with the mass profiles reported in the top panel. Green lines as in Fig.~2.}
\end{center}
\end{figure*}


\begin{references}

\reference{}Akamatsu, H., Hoshino, A., Ishisaki, Y., Ohashi, T., Sato, K.,
Takei, Y., \& Ota, N. 2011, PASJ, 63, 1019

\reference{}Arnaud, M., Pointecouteau, E., \& Pratt, G. W. 2005, A\&A, 441,
893

\reference{}Arnaud, M., Pointecouteau, E., \& Pratt, G. W. 2007, A\&A, 474,
L37

\reference{}Battaglia, N., Bond, J. R., Pfrommer, C., \& Sievers, J. L. 2013,
ApJ, 777, 123

\reference{}Battaglia, N., Bond, J. R., Pfrommer, C., \& Sievers, J. L. 2014,
ApJ, submitted [preprint arXiv:1405.3346]

\reference{}Bautz, M.W., Millwe, E.D., \& Sanders, J. S. et al. 2009, PASJ,
61, 1117

\reference{}Bonamente, M., Landry, M., Maughan, B., Giles, P., Joy, M., \&
Nevalainen, J., 2013, MNRAS, 428, 2812

\reference{}Brunetti, G., \& Lazarian, A. 2011, MNRAS, 410, 127

\reference{}Burns, J. O., Skillman, S., W., \& O'Shea, B., W. 2010, ApJ, 721,
1105

\reference{}Cavaliere, A., \& Lapi, A. 2013, Phys. Rep., 533, 69

\reference{}Cavaliere, A., Lapi, A., \& Fusco-Femiano, R. 2011a, ApJ, 742, 19

\reference{}Cavaliere, A., Lapi, A., \& Fusco-Femiano, R. 2011b, A\&A, 525,
110

\reference{}Cavaliere, A., Lapi, A., \& Fusco-Femiano, R. 2009, ApJ, 698, 580

\reference{}Eckert, D. et al. 2012, A\&A, 541 57

\reference{}Eckert, D., Molendi, S., Vazza, F., Ettori, S., \& Paltani, S.
2013, 551, A22

\reference{}Fabian, A. C. 2012, ARA\&A, 50, 455

\reference{}Fujita, Y., Ohira, Y., \& Yamazaki, R. 2013, ApJ, 767, L4

\reference{}Fusco-Femiano, R., Cavaliere, A., \& Lapi, A. 2009, ApJ, 705,
1019

\reference{}Fusco-Femiano, R., \& Lapi, A. 2013, ApJ, 771, 102

\reference{}Fusco-Femiano, R., \& Lapi, A. 2014, ApJ, 783, 76

\reference{}Gonzalez, A. H., Zaritsky, D., \& Zabludoff, A. I. 2007, ApJ,
666, 147

\reference{}Hinshaw, G., Larson, D., Komatsu, E., et al. 2013, ApjS, 208, 19

\reference{}Hoshino, A., Henry, P. H., Sato, K., et al. 2010, PASJ, 62, 371

\reference{}Ichikawa, K., Matsushita, K., Okabe, N. et al. 2013, ApJ, 766, 90

\reference{}Inogamov, N.A., \& Sunyaev, R.A. 2003, Astron. Lett., 29, 791

\reference{}Kawaharada, M., Okabe, N., Umetsu, K., et al. 2010, ApJ, 714, 423

\reference{}Kolmogorov, A. 1941, Dokl. Akad. Nauk SSSR, 30, 301

\reference{}Kravtsov, A.V., \& Borgani, S. 2012, ARA\&A 50, 353

\reference{}Lapi, A., Fusco-Femiano, R., \& Cavaliere, A. 2010, A\&A, 516, 34

\reference{}Lapi, A., Cavaliere, A., \& Menci, N. 2005, ApJ, 619, 60

\reference{}Lau, E. T., Kravtsov, A. V., \& Nagai, D. 2009, ApJ, 705, 1129

\reference{}Mahdavi, A., Hoekstra, H., Babul, A., \& Henry, J.P. 2008, MNRAS,
384, 1567

\reference{}Mahdavi, A., Hoekstra, H., Babul, A., Bildfell, C., Jeltema, T.,
\& Henry, J.P. 2013, ApJ, 767, 116

\reference{}Mathiesen, B., Evrard, A. E., \& Mohr, J. J. 1999, ApJ, 520, L21

\reference{}Monin, A.S., \& Yaglom, A.M. 1965, \emph{Statistical
Hydromechanics} (Moscow: Nauka)

\reference{}Morandi, A., \& Cui, W. 2014, MNRAS, 437, 1909

\reference{}Nagai, D., Vikhlinin, A., \& Kravtsov, A. V. 2007, ApJ, 665, 98

\reference{}Nagai, D., \& Lau, E. T. 2011, ApJL, 731, 10

\reference{}Nelson, K., Lau, E.T., Nagai, D., Rudd, D.M. \& Yu, L. 2014, ApJ,
782, 107

\reference{}Okabe, N., Takada, M., Umetsu, K., Futamase, T., \& Smith, G.P.
2010, PASJ, 62, 811

\reference{}Okabe, N., Smith, G.P., Umetsu, K., Takada, M., \& Futamase, T.
2013, ApJ. 769, L35

\reference{}Okabe, N., Umetsu, K., Tamura, T., et al. 2014, PASJ, 66, 99

\reference{}Panagoulia, E.K., Fabian, A.C., \& Sanders, J.S. 2014, MNRAS,
438, 2341

\reference{}\textsl{Planck} Collaboration Ade, P. A. R., Aghanim, N.,
Armitagr-Caplan, C., Arnaud, M., et al. 2014, A\&A, 556, 54

\reference{}\textsl{Planck} Collaboration Ade, P. A. R., Aghanim, N., Arnaud,
M., Ashdown, M., et al. 2013, A\&A, 550, 131

\reference{}Petrosian, V., \& East, W.E. 2008, ApJ, 682, 175

\reference{}Pratt, G.W., Arnaud, M., Piffaretti, R., Bohringer, H., Ponman,
T.J., Croston, J.H., Voit, G.M., Borgani, S., \& Bower, R.G. 2010, A\&A, 511,
A85

\reference{}Rasia, E. et al. 2012, New Journal of Physics, 14, 055018

\reference{}Reiprich, T. H., Basu, K., Ettori, S., Israel, H., Lovisari, L.,
Molendi, S., Pointecouteau, E., \& Roncarelli, M. 2013, Sp. Sci. Rev., 177,
195

\reference{}Roncarelli, M., Ettori, S., Borgani, S., Dolag, K., Fabjan, D.,
\& Moscardini, L. 2013, MNRAS, 432, 3030

\reference{}Rossetti, M. \& Molendi, S. 2010, A\&A, 510, 83

\reference{}Sato, K., Matsushita, K., Yamasaki, N.Y., Sasaki, S., \& Ohashi,
T. 2014, PASJ, 66, 85

\reference{}Shaw, L. D., Nagai, D., Bhattacharya, S., \& Lau, E. T. 2010,
ApJ, 725, 1452

\reference{}Simionescu, A. et al. 2011, Science, 331, 1576

\reference{}Simionescu, A., Werner, N., Urban, O., Allen, S.W., Fabian, A.C.,
Mantz, A., Matsushita, K., Nulsen, P.E.J., Sanders, J. S., Sasaki, T., et al.
2013, ApJ, 775, 4

\reference{}Tozzi, P., \& Norman, C. 2001, ApJ, 546, 63

\reference{}Vazza, F., Brunetti, G., Gheller, C., Brunino, R., \& Bruggen, M.
2011, A\&A, 529, A17

\reference{}Vazza, F., Eckert, D., Simionescu, A., Bruggen, M., \& Ettori, S.
2013, MNRAS, 429, 799

\reference{}Vikhlinin, A., Kravtsov, A.V., Burenin, R.A., et al. 2009a, ApJ,
692, 1060

\reference{}Vikhlinin, A., et al. 2009b, ApJ, 692, 1033

\reference{}Voit, G. M. 2005, Rev. Mod. Phys., 77, 207

\reference{}Walker, S., Fabian, A., Sanders, J., George, M., \& Tawara, Y.
2012, MNRAS, 422, 3503

\reference{}Walker, S., Fabian, A., Sanders, J.S., Simionescu, A, \& Tawara,
Y. 2013, 432, 554

\reference{}Wang, Q.D., \& Walker, S. 2014, MNRAS, 439, 1796

\reference{}Zhuravleva, I., Kravtsov, A., Lau, E.T., Nagai, D., \& Sunyaev,
R. 2013, MNRAS, 428, 3274

\end{references}
\end{document}